\documentclass[12pt]{article}
\usepackage{graphicx}

\begin{document}

\title{On accuracy assessment of celestial reference frame VLBI realizations}
\author{Zinovy Malkin \\ Central (Pulkovo) Astronomical Observatory RAS, \\
Pulkovskoe~ch. 65-1, St.~Petersburg 196140, Russia \\ malkin@gao.spb.ru}
\date{March 04, 2007}
\maketitle

\begin{abstract}
In this paper, we propose to use the scatter of celestial pole offset (CPO)
series obtained from VLBI observations as a measure of the accuracy
of the celestial reference frame (CRF) realizations.
Several scatter indices (SIs) including those proposed for the first time
are investigated.
The first SI is based on analysis of residuals of CPO series
w.r.t. Free Core Nutation (FCN) model.
The second group of SIs includes Allan deviation and its extensions,
which allow one to treat unequal and multidimensional observations.
Application of these criteria to several radio source catalogues showed
their ability to perform a preliminary assessment of the quality of
radio source catalogues, and 2D Allan deviation estimate seems to be a most
sensitive measure.
However, in common case, the sensitivity of tested criteria is yet not
sufficient to distinguish between radio source catalogues of the highest
quality.
Proposed extensions of Allan deviation, weighted and multidimensional,
can be effectively used also for statistical analysis of astro-geodetic
and other time series.
\end{abstract}


\section{Introduction}
\label{intro}

Very long baseline interferometry (VLBI) is the base technique
provided International Celestial Reference Fra\-me (ICRF)
realized as a set of radio source coordinates (Ma et al. 1998).
Improvement of the ICRF accuracy is one of the primary task for
International VLBI Service for Astrometry and Geodesy (IVS),
and assessment of the ICRF accuracy is of primary importance
in its improvement (Schl\"uter et al. 2002).

Many studies of the accuracy of the ICRF have been performed during last years.
Some authors have investigated the time behavior of radio source position, e.g.
Gontier et al. (2001), Feissel-Vernier (2003).
Others have studied the accuracy of different CRF realizations by means of
investigation of stability of the coordinate axis,
e.g. Arias (1988), Arias (2004).
However, existing method allow us to investigate only {\it differences} between
Celestial Reference Frame (CRF) realizations, or, in other words, radio
source position catalogues.
There is no evident and received method to assess the
{\it absolute} ICRF accuracy.

In this paper, we consider possible criteria, which allow one to assess
the accuracy of radio source catalogues from its impact on results of
determination of the celestial pole offset (CPO) from VLBI observations.
We propose to use the scatter of CPO
time series obtained from VLBI observations as a measure of the accuracy
of the CRF realizations.
Several scatter indices (SIs), including those proposed for the first time
are investigated.
The first SI is based on analysis of residuals of CPO series
from a reference series obtained as the IAU2000A precession-nutation model
with addition of the Free Core Nutation (FCN) contribution.
The second group of SIs includes Allan deviation (ADEV) and its extensions.
Two such extensions are proposed. The first one, weighted Alan variance
WADEV, which allow one to treat unequal measurements, which we mostly deal with
in geodesy. The second extension of the classical definition is
weighted multidimensional Allan deviation estimator WMADEV, which can be
applied to combined analysis of several associated time series.

We have applied these SIs to several radio source position catalogues.
The tested catalogues of the first group were obtained
in the framework of the joint pilot project of the International Earth
Rotation and Reference Systems Service (IERS) and the IVS
(Ma 2004, Call for Participation).
The second group of catalogues consists of
the latest submissions of IVS Analysis Centers to the IVS.
The latest ICRF realization, ICRF-Ext.2 (Fey 2004), was also tested
in this study.
Based on the results of this work, we can make a conclusion on ability
of various SIs to detect difference in accuracy (quality) of the CRF
realizations provided by VLBI Analysis Centers.


\section{Scatter indices}

Investigation of the scatter of a time series of geodetic, geodynamical
or other data under interest is proven to be a powerful tool
for assessment of its quality and statistical characteristics.
The scatter of the time series can be computed in different way.
We will use the term "scatter index" (SI) to distinguish the results
obtained by different methods.
Advantage of using SIs for investigation of time series is its independence,
in the most of cases, of systematic errors in tested results, trends and
low-frequency variations.

\subsection{Residuals in CPO series}
\label{CPO}

Both precision and accuracy of the CPO
estimates directly depend on accuracy of radio source positions
used during VLBI data processing.
There is no external {\it absolutely accurate} reference series that may
be used for comparison,
for only VLBI technique provides the highly-accurate CPO measurements.

The scatter of Earth Orientation Parameters (EOP) time series is widely used
for many years for comparison of EOP results obtained in different analysis
centers.
We will try to apply it to comparison of CRF realization.
Similar test was used in Feissel-Vernier et al. (2006), where the authors
compared several radio source catalogues obtained in the course of their
work using
the WRMS differences with the IAU2000 precession-nutation model.
However, it seems that such a test can be improved by using of the observed
correction to the IAU2000 model, the contribution of the Free Core Nutation
in the first place.
In such a case, the residuals between VLBI nutation series and model
becomes much less, and the test, hopefully, will be more sensitive.
Thus we used the IAU2000A model with addition of the FCN contribution as
the reference series.
For our purpose, we computed the FCN contribution by smoothing
of the differences between the IVS combined EOP series and the IAU2000A model.

However, this SI cannot be considered as really independent estimate of
the CPO scatter since any FCN model derived from a comparison of
observations with theory depends on results of VLBI data processing.
Evidently, such an estimate implicitly depends on underlying CRF realizations.
Hence, FCN criterion provides only comparison
of tested radio source catalogue realization with some averaged CRF
realization.


\subsection{Allan deviation}
\label{ADEV}

Allan variance estimator was originally developed for investigation of
noise parameters of frequency standards.
In the last years, Allan-variance-based techniques is often applied to
analysis of geodetic and geodynamical time series.
Usually, square root of Allan variance ADEV is used for this purpose.
IERS EOP Product Center employs this technique for assessment of the
statistical characteristics of the EOP series (Gambis 2002).
ADEV estimates are used for investigation of stochastic properties
of station coordinates time series, e.g. Malkin \& Voinov (2001),
Le Bail \& Feissel-Vernier (2003), Feissel-Vernier \& Le Bail (2005),
Le Bail (2006).
Feissel-Vernier (2003) has also used Allan variance for
analysis of the noise in radio source position time series.

Allan deviation estimate for time series $y_1,y_2,\dots,y_n$
is given by
\begin{equation}
\label{eq:adev}
ADEV = \sqrt{\frac{1}{2(n-1)}\sum_{i=1}^{n-1}(y_{i+1}-y_i)^2}\,.
\end{equation}

Commonly speaking, this formula is valid for equal measurements.
However, in geodetic and astrometric practice, we usually deal with
unequal measurements.
In this case, we have the measurements with associated uncertainties
$s_1,s_2,\dots,s_n$.
To treat unequal data, the following extension of ADEV can be proposed.
\begin{eqnarray}
\label{eq:wadev}
WADEV = \sqrt{\frac{1}{2p}\sum_{i=1}^{n-1}p_i(y_{i+1}-y_i)^2}\,, \\
p=\sum_{i=1}^{n-1}{p_i}\,, \quad p_i=\frac{1}{s_i^2+s_{i+1}^2} \,. \nonumber
\end{eqnarray}

\begin{figure}
\centering
\resizebox{\hsize}{!}{\includegraphics[clip]{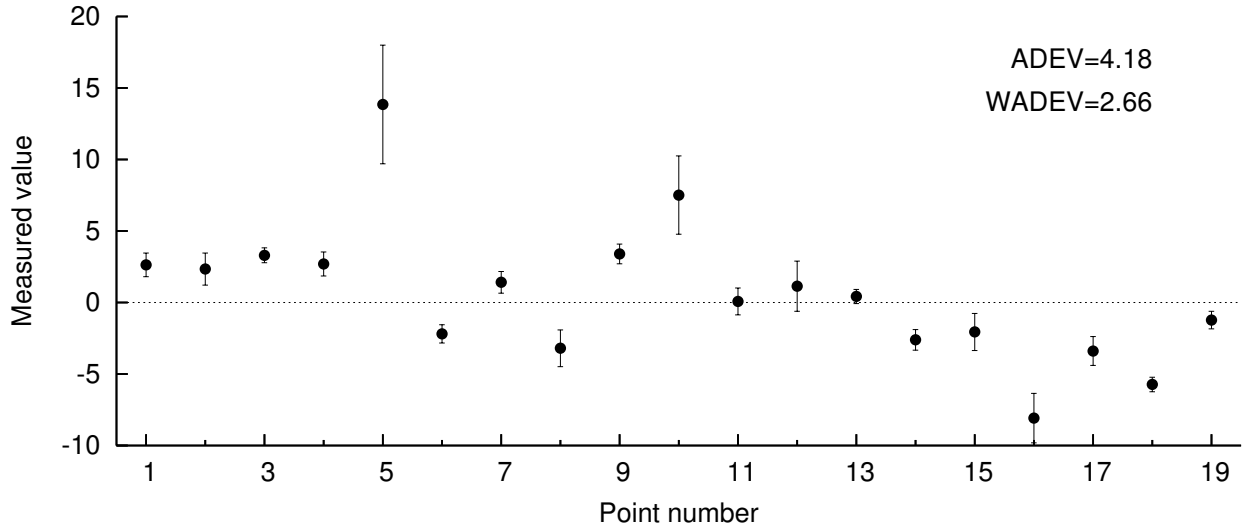}}
\caption{An example of estimation of classical (ADEV) and weighted (WADEV)
Allan deviation for unequal time series.}
\label{fig:WADEV_example}
\end{figure}

Fig.~\ref{fig:WADEV_example} shows an example of actual series of station
coordinates, which show that the difference between ADEV and WADEV values
may be quite substantial.

Further extension of WADEV estimator is useful for analysis of some
types of time series.
When the statistical parameters of astro-ge\-o\-de\-tic time series are
investigated, we, in fact,
often deal with multidimensional values, e.g. terrestrial coordinates and/or
velocity (3D or 6D),
celestial coordinates and/or proper motion (2D or 4D), and so on.
In such a case, we can use multidimensional notation
$y_i=(y_i^1,y_i^2,\dots,y_i^k)$
with the associated k-dimensional uncertainties
$s_i=(s_i^1,s_i^2,\dots,s_i^k)$.
Thus, we can define the k-dimensional Allan deviation as

\begin{eqnarray}
\label{eq:WMADEV}
WMADEV = \sqrt{\frac{1}{2p}\sum_{i=1}^{n-1}p_i|y_{i+1}-y_i|^2}\,, \\
p=\sum_{i=1}^{n-1}{p_i}\,, \quad
p_i=\frac{1}{\sum\limits_{j=1}^k [(s_i^j)^2 + (s_{i+1}^j)^2]} \,. \nonumber
\end{eqnarray}

It is needless to say that both WADEV and WMADEV estimates can be easily
generalized, like classical ADEV estimate, for different sampling interval
as well as overlapping intervals.


\section{Comparison of catalogues}
\label{Comparison}

For our study, we computed several CPO time series
with the same processing options, except different underlying radio
source catalogues.  We used 504 R1 and R4 VLBI sessions
observed in the period 2002.0--2007.0.
Then we compute the SI described above for all the series.

We included in this comparison three groups of VLBI CRF realizations.
First, we used eight radio source catalogues obtained
in the framework of the joint IERS-IVS Pilot Project mentioned above.
The catalogues were provided by eight IVS Analysis Centers:
AUS (Geoscience Australia),
BKG (Bundesamt f\"ur Kartographie und Ge\-o\-d\"asie, Germany),
DGFI (Deu\-t\-s\-c\-hes Ge\-o\-d\"a\-ti\-sches For\-sch\-ungsInstitut,
Germany),
GSF (NASA Goddard Space Flight Center, USA), JPL (Caltech/NASA Jet Propulsion
Laboratory, USA), MAO (Main Astronomical Observatory of National Academy
of Sciences of Uk\-ra\-i\-ne),
SHA (Shanghai Astronomical Observatory, China),
USN (U.~S.~Naval Observatory, USA).
Each Analysis Center has submitted two version of catalogue.  The first one
was computed using all available VLBI data. For the second version, only
the observations made since 1990 have been used.
Then we included in our study six radio source catalogues
submitted by IVS Analysis Centers in 2006--2007 in the framework of
routine operations.
The latest ICRF realization,
ICRF-Ext.2 (Fey 2004), was also used in our study.

Results of this test are presented in Table~\ref{tab:comparison}.
In the table, FCN column shows the WRMS differences between CPO
series computed with given radio source catalogue and IAU2000A+FCN
reference series, ADEV and WADEV columns contain classical and weighted
Allan variance estimates.
The values presented in the FCN, ADEV and WADEV columns
are computed as the mean of the estimates for $X$ and $Y$ celestial
pole coordinates.
The WMADEV column contains 2D estimates computed with (\ref{eq:WMADEV}),
$k=2$, for both pole coordinates.

\begin{table}
\centering
\caption{Scatter indices for celestial pole offset series obtained with
different radio source catalogues. See text for details. Unit: $\mu$as.}
\label{tab:comparison}
\begin{tabular}{lcccc}
\hline
Catalogue  & FCN & ADEV & WADEV & WMADEV \\
\hline
\multicolumn{5}{c}{Pilot Project catalogues, 1979--2005} \\
AUS1       & ~99 & 116 & 108 & 153 \\
BKG1       & ~96 & 111 & 104 & 147 \\
DGF1       & ~96 & 115 & 104 & 148 \\
GSF1       & ~96 & 111 & 103 & 146 \\
JPL1       & ~98 & 113 & 105 & 149 \\
MAO1       & ~96 & 113 & 104 & 147 \\
SHA1       & ~96 & 114 & 104 & 148 \\
USN1       & ~96 & 114 & 104 & 147 \\
\multicolumn{5}{c}{Pilot Project catalogues, 1990--2005} \\
AUS2       & ~97 & 113 & 104 & 148 \\
BKG2       & ~95 & 112 & 103 & 146 \\
DGF2       & ~96 & 111 & 103 & 146 \\
GSF2       & ~95 & 111 & 103 & 146 \\
JPL2       & ~96 & 110 & 102 & 145 \\
MAO2       & ~95 & 111 & 103 & 145 \\
SHA2       & ~96 & 112 & 104 & 147 \\
USN2       & ~95 & 111 & 102 & 145 \\
\multicolumn{5}{c}{Latest catalogues} \\
AUS2006B   & ~97 & 112 & 103 & 147 \\
BKG2006C   & ~96 & 112 & 103 & 146 \\
CGS2006A   & ~95 & 110 & 102 & 144 \\
GSF2007A   & ~96 & 111 & 103 & 146 \\
IAA2006A   & ~97 & 114 & 106 & 150 \\
USN2006C   & ~95 & 110 & 102 & 144 \\
\hline
ICRF-Ext.2 & ~99 & 120 & 111 & 157 \\
\hline
\end{tabular}
\end{table}

One can see that the accuracy of all the compared catalogues as estimated
by this method is close to each other.
Nevertheless, some discrepancies can be clearly seen.
In particular, the ICRF-Ext.2 showed the worst result,
maybe because ITRF-Ext.2 212 {\it defining} source positions were saved
from the first ICRF.
This can be mentioned that, in the analyzed data,
186 {\it defining} sources were observed of total number of 507 ({\it i.e.} 36.6\%),
and there were 307155 observations of {\it defining} sources of total number
of 953637 ({\it i.e.} 32.2\%).
If so, it might be inadvisable to keep coordinates
of the {\it defining} ICRF sources in the successive ICRF updates.

The organizers of the IERS-IVS Pilot Project requested from the
participating Analysis Centers to submit two versions of catalogues
obtained using all available data (version 1) and data from 1900 only
(version2). The intention was to investigate an impact of data selection
on the quality of CRF realization.
From our test, one can see that catalogues of version~2 show small but
steady improvement of the accuracy.
This effect can be also explained by the greater weight of
R1/R4 sessions in the catalogues of version~2 w.r.t. catalogues of
the version~1 though.
On the other hand, previous studies (Malkin 2004a,b) have shown that both
EOP and precession parameters derived from VLBI observations show much
less uncertainty when only data from 1990 are used for analysis.

As to comparison of different SIs considered in this test,
2D Allan variance seems to be the most sensitive criterion.
It is advisable to use the weighted multidimensional
Allan deviation for investigation of other astro-geodetic
multidimensional time series.


\section{Conclusion}
\label{concl}

In this paper, we tested a possibility of using the scatter of celestial
pole offset (CPO) series obtained from VLBI observations as a measure of the
accuracy of the CRF realizations.
Several scatter indices (SIs), including those proposed for the first time
are investigated.

The first SI is based on analysis of residuals of CPO series
w.r.t. IAU2000A precession-nutation model supplemented with the
FCN contribution. It turned out to be less sensitive than
other tested SIs.  Besides this SI cannot be considered as fully
independent since both IAU2000A model and FCN contribution depend
on previous VLBI data analysis, in particular, used CRF realizations.
Conversely, the SIs based on Allan deviation and its extensions
proposed in this paper provide independent estimate of the quality
of CRF realizations.

Application of these criteria to several radio source catalogues has shown
their ability to perform a preliminary assessment of the quality of
radio source catalogues, and 2D Allan deviation estimate seems to be a most
sensitive measure.
However, the sensitivity of tested criteria is yet not sufficient
to distinguish between radio source catalogues of the highest quality.

Proposed extensions of the Allan deviation estimator, weighted WADEV
and multidimensional WMADEV, can be effectively used also for statistical
analysis of geodetic
and other time series, {\it e.g.} Earth Orientation Parameters and station
position time series.

\end{document}